# Towards a Dutch hybrid quantum/HPC infrastructure

Olaf Schüsler
QuTech
*Delft, The Netherlands*
o.m.schusler@tudelft.nl

Ariana Torres-Knoop
SURF B.V.
*Utrecht, The Netherlands*
ariana.torres@surf.nl

Jaap Dijkshoorn
SURF B.V.
*Utrecht, The Netherlands*
jaap.dijkshoorn@surf.nl

Christiaan Hollemans
QuTech
*Delft, The Netherlands*
christiaan.hollemans@tno.nl

Bas van der Vlies
SURF B.V.
*Utrecht, The Netherlands*
bas.vandervlies@surf.nl

Richard Versluis
QuTech
*Delft, The Netherlands*
richard.versluis@tno.n

*Abstract*—Quantum Inspire has taken important steps to enable quantum applications by developing a setting that allows the execution of hybrid algorithms. Currently, the setting uses a classical server (HPC node) co-located with the quantum computer for the high frequency coupling needed by hybrid algorithms. A fast task manager (dispatcher) has been developed to orchestrate the interaction between the server and the quantum computer. Although successful, the setting imposes a specific hybrid job-structure. This is most likely always going to be the case and we are currently discussing how to make sure this does not hamper the uptake of the setting. Furthermore, first steps have been taken towards the integration with the Dutch National High-Performance Computing (HPC) Center, hosted by SURF. As a first approach we have setup a setting consisting of two SLURM clusters, one in the HPC (C1) and the second (C2) co-located with Quantum Inspire API. Jobs are submitted from C1 to C2. Quantum Inspire can then schedule with C2 the jobs to the quantum computer. With this setting, we enable control from both SURF and Quantum Inspire on the jobs being executed. By using C1 for the jobs submission we remove the accounting burden from Quantum Inspire. By having C2 co-located with Quantum Inspire API, we make the setting more resilient towards network failures. This setting can be extended for other HPC centers to submit jobs to Quantum Inspire backends.

## I. Introduction

Quantum computers are devices that process information by taking advantage of the quantum-mechanical properties of their building blocks, the qubits. By doing so, they can harness work in a powerful and efficient way and perform certain operations with an exponential speed-up. There are many fields that could benefit from such a speed-up, for example machine learning, financial modeling, logistic optimization, climate simulations, etc. [1]. Notably quantum computers are expected to excel at simulating quantum systems, like the ones present in chemistry and material science [2].

Harnessing the power (time and energy-to-solution) of quantum computers in relevant use cases and applications is not trivial. Firstly, it requires the research and development of algorithms that can leverage the fundamentally different quantum technology. Secondly, it requires the development of a full stack to implement and execute the quantum algorithms in the quantum hardware. The full stack needs to be integrated to the existing classical ecosystem. Thirdly, the implementation and integration needs to be tested and optimized. An optimization towards global usage might be complicated due to the different hardware. Finally, the development of quantum applications requires the exploration and translation of potential use cases from a classical execution into a quantum-classical (hybrid) execution. The user community will have to learn to rethink their problems [3].

It is generally expected that quantum computers will be used as accelerators for classically complex computational tasks. The classical host will execute the main application and off-load some subroutines or tasks to the quantum computer. In some cases, quantum computers will also benefit from on-loading tasks into classical resources, for example for Quantum Error Correction (QEC) [4,5]. Moreover, as quantum computers improve, the need to on-load tasks to classical resources likely will only grow.

An optimal implementation of the hybrid quantum-classical applications and the integration of the quantum and classical resources is critical to ensure the quantum advantage obtained from the algorithms is not killed. Co-design and co-development of the applications with the emerging quantum infrastructure and the existing classical infrastructure is fundamental. Furthermore, quantum applications require the orchestration of quantum and classical programs, often based on different languages, data formats or invocation mechanisms [6,7].

When considering the integration and interfaces between the classical and quantum resources (hybrid infrastructure), many different strategies and scenarios are possible [8]. From a functional perspective, the hybrid infrastructure should allow for easy application development. From an operational perspective, the hybrid infrastructure should optimize the time, energy and science-to-solution.

## II. Hybrid infrastructure: operational perspective

To optimally execute hybrid quantum-classical applications we need to identify when and how classical resources play a role in the application. In this paper we will refer to three main operational levels: application, task and control (see Figure 1).

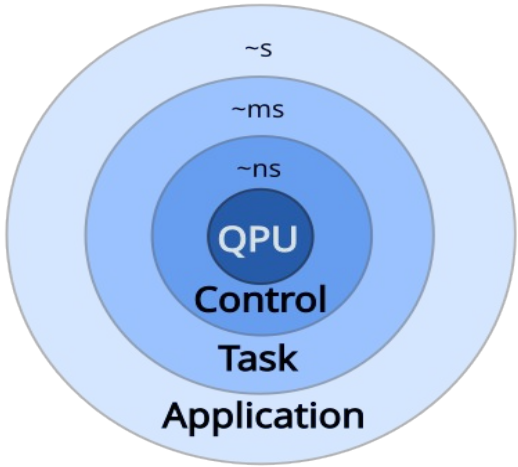

Fig. 1. In the execution of hybrid application we can distinguish three main operational layers: application, task, control. They all have different requirements, in particular related with the latency and data exchange.

*1. Application level:* Most hybird applications will be part of complex workflows executed in a main classical host. Only some tasks will be off-loaded to the quantum computer. The execution of the full application will require the orchestration of classical and quantum tasks. To ensure the quantum tasks do not slow down the high-level application, feedback latency of the order of seconds is desirable.

*2. Task level:* A quantum task might require on-loading of some sub-tasks to classical resources, for example, for quantum circuit cutting [9], circuit architecture search [10], classical optimization in variational algorithms [11], etc. To ensure the quantum task is not slowed down due to the on-loading to classical resources, feedback latency of the order of the quantum measurements are desirable. Although very variable and hardware dependant, in general a latency of µs – ms should be expected. A lower feedback latency might be neccesary to enable the execution of protocols that require e.g. mid-circuit measurements.

*3. Control level*: The quantum control hardware executes gates and measurements in different manners depending on the quantum-chip technology. It generally also stores, reads and interprets the received instructions in order to generate respective pulse sequences [12]. To enable any useful control of a qubit, programmable control flow that operates on the timescale of nanoseconds is a requirement [13].

Each operational level manages computational resources differently. For the application level, assuming the main classical host is a High Performance Computing (HPC) center, common examples of job managers are SLURM [14] and OpenPBS [15]. New developments on scheduling are done with tools like Flux [16], Hyperqueue [17] and QCGpilot [18] due to their capability to run efficiently and easily on modern heterogeneous supercomputers.

For the task level, the resource manager needs to be able to work in the time scale of the QPU or lower. Although some existing job managers could be adapted to handle these timescales, the approach so far has been to develop job managers that take into account the specific needs of the quantum applications. The classical sub-tasks are executed in a classical runtime (see Figure 2). A classical runtime could be the main classical host (e.g. HPC Center or user laptop) or for example a small server in the proximity of the quantum computer. This last setup is often referred to as "hybrid runtime" and is for example used by Qiskit Runtime [19] and Amazon Braket Hybrid Jobs [20]. The optimization of the communication between computational resources strongly depends on the hybrid infrastructure set-up (stand-alone, co-location, distributed) [7].

At the control level, to perform logical operations, the hardware specific instructions need to be very carefully timed by the classical control electronics. This is often achieved using a field programmable gate array (FPGA) [21].

Due to the scarcity of quantum resources, an additional "job manager" is in some cases needed to orchestrate the quantum tasks arriving from different sources, for example, different users in an HPC center, different HPC centers or in some cases individual users (see Figure 2). Queue-based access might not be the most suitable when executing tasks that require a high frequency coupling between the main classical host and the quantum computer. Instead, access to the quantum computer might have to be reserved for the duration of the task. Such a model has already been used by for example IBM Q [22].

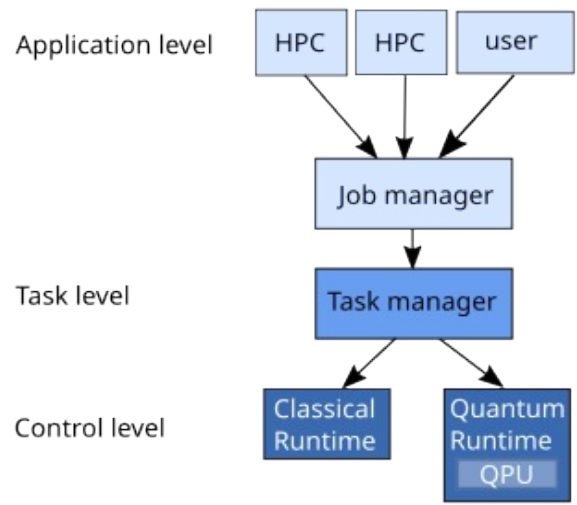

Fig. 2. The high-level hybrid applications are run in a main classical host (HPC or user). Subroutines of quantum tasks are off-loaded to the quantum backend. In the quantum backend, different tasks are managed by a job manager to ensure the quantum computer runs at its maximum capacity while providing the necessary resources for a given application. When the quantum task requires classical resources, the task manager orchestrates the classical and quantum runtime. The classical runtime can be an independent server or the main classical host (e.g. HPC center itself).

## III. Quantum Inspire

Quantum Inspire [23] is Europe's first public quantum computing platform. The platform focuses primarily on training, education, and the development of applications, so that more people can use the quantum computer as it develops further and becomes more widely available. It consists of a number of layers including quantum hardware (a processor made of semiconductor 'spin qubits' and a processor made of superconducting transmon qubits, classical control electronics, and a software front-end with a cloud-accessible web-interface (full-stack)).

Quantum Inspire's programming language, in which quantum algorithms are written and executed, is called cQASM [24] (similar to the Open Quantum Assembly Language [25]). Algorithms can be programmed using the web-editor or using the SDK. This SDK provides a thin layer between the QI application programming interface (API) and other programming platforms using Python [26, 27].

Recently, much effort has been done in the development of Quantum Inspire to enable the execution of hybrid tasks. Together with SURF [28], the IT collaboration for education and research in the Netherlands and host of the Dutch National supercomputer Snellius, preliminary work has also been done to enable the execution of hybrid applications. In the following section we will describe the current hybrid setting and our experience so far.

## IV. QUANTUM INSPIRE FOR HYBRID EXECUTIONS

For the hybrid infrastructure of Quantum Inspire and SURF we have taken a distributed approach [7]. In this approach, the main classical host (HPC or individual user) off-loads a quantum task to the quantum platform. The quantum task can be a "pure quantum" task or a hybrid task.

The hybrid task is executed in the classical and quantum runtimes. The set of supported programming languages for hybrid tasks is typically restricted. Quantum Inspire currently accepts Python code.

### A. Quantum Inspire task manager (dispatcher)

The Quantum Inspire backend consists of a classical runtime and several quantum runtimes. The quantum runtimes can either be a hardware backend (QPU) or an emulator [29]. Communication with the various runtimes happens via ZeroMQ [30]. Since switching between the various runtimes needs to happen as quickly as possible, the request/reply pattern is employed [31]. Whenever a request for execution arrives at the task manager, these requests are propagated to the necessary runtimes. Replies can be returned directly on completion of the classical or quantum task and trigger the next step in the execution. This communication scheme guarantees the minimal wait time per component.

When executing pure quantum tasks (a quantum circuit) the task manager sends the circuit directly to the requested quantum runtime (see Figure 2). For hybrid tasks, the task manager takes care of the lifecycle management. As soon as it determines the job is a hybrid algorithm, it requests the classical runtime to execute the python script and return a quantum circuit. This circuit is forwarded to the quantum runtime, where it is executed. This continues to run, until a stop condition is met. This can either be a timeout, or a user defined end state. A simple "ping-pong" implementation between the task manager (dispatcher) and the runtimes is presented in Appendix A.

Timing information of the preliminary implementation is given in Table 1. The initialization step spins up the runtime only once on every hybrid task. This start-up time however, can be shortened by hot loading environments. For the execution step, no computations were performed to give raw timing information.

TABLE I. TIMING INFORMATION OF THE TASK MANAGER

| Step | Runtime |
|---|---|
| Initialization | 3 s 792 µs |
| Execution | 17 µs |
| Termination | 200 µs |

As presented in Table 1, once the runtime is spun up, the developed task manager allows the orchestration between the classical and quantum runtimes within the µs tiemscale. This is in line with the timescale of a task.

### B. Quantum Inspire job manager

The Quantum Inspire system contains a ReST API. This API gives users access system information and their own content. Users can, for example, get the status of the various backends. At the time of writing this is limited to whether the backend is executing, calibrating, offline or idle. Real time queueing information is not (yet) available. The user can also allow the API to handle their own content. Projects and algorithms can be created, but also jobs submitted or compiled algorithms and results fetched. This can be done via any of the offered end user interfaces, like the web frontend or the SDK. Furthermore, users can also communicate directly with the API.

Any job submitted via this API ends up in a queue and is subject to scheduling. As a first approach, jobs are scheduled naively, via a FIFO queue. Integration with HPC systems however will prompt more elaborate systems like priority queues or reservations. Different scenarios and possibilitites are currently under consideration.

### C. Application manager

In HPC centers, resource managers like SLURM of OpenPBS are used. These managers oversee prioritizing jobs, checking resources, and launching jobs. Quantum applications cause the management to be more constrained due to, among other, the scarcity of resources, the time-bound execution of an algorithm and the asymmetry of resources [3].

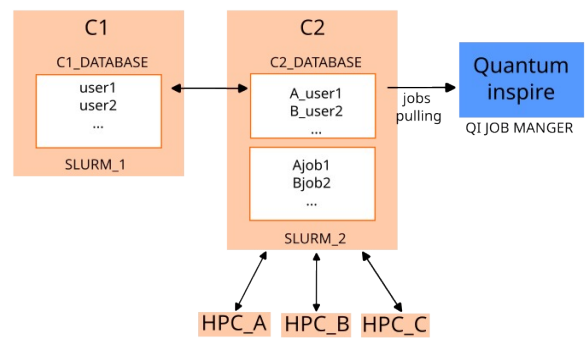

Fig. 3. Schematic representation of the 2-SLURM clusters setup for hybrid quantum-classical applications. The setup can in principle be used to coordinate the access of several HPC centers to one quantum backend. It enables control from both the HPC centers and the quantum resources.

For our first approach to a hybrid Quantum Inspire / HPC (SURF) infrastructure, we will use the HPC as the main classical host and a small server co-located with the quantum computer as the classical runtime (see Figure 2). SLURM is used as the application manager and the Quantum Inspire dispatcher as the task manager. The main challenge for integration in this case is the proper submission of jobs from the HPC center to the quantum backend and the efficient job handling in the quantum backend. In a HPC cluster normally only one resource manager is in control of all hardware in the cluster. In this approach, we need to orchestrate the pipeline with the knowledge that in fact two resource managers are in charge within one pipeline.

The integration set-up consists of "2-SLURM" clusters (see Figure 3). The first SLURM cluster (C1) is the HPC center and the second SLURM cluster (C2) is co-located with the Quantum Inspire API. Jobs are submitted from the C1 to C2. A batchjob is created with the payload needed for Quantum Inspire. The payload is json formatted and then submitted from C1 with CURL to the REST-API of the SLURM scheduler [32] at C2. A simple payload example and submission code are presented in Appendix B. The submit code can be easily included in any batch job.

The C2 cluster will always accept jobs. These jobs are queued and waiting for execution. When the C2 Cluster is enabled (by Quantum Inspire), jobs will be run on the quantum computer.

In this approach, the HPC center and quantum resources are managed independently. This can help the performance of largely imbalanced application with low-coupling frequency between the main classical host and the quantum backend. For high-frequency coupling, this approach might get increasingly complex.

In a first instance, users can retrieve their data directly for the Quantum Inspire data base. Another approach is to stage data securely on a cloud space for example SURFDrive [33] or any other webdav compliant cloud storage.

Other HPC centers could talk to C2 in the same way. On C2 a SLURM-USER-TOKEN has to be created for all participating HPC centers (one each). Jobs submitted to C2 are being accounted in de SLURM Database (see Figure 3).

## V. CHALLENGES AND LEARNINGS

One of the main drawbacks of this design is the fact that users cannot get access to the classical runtime where the quantum computer is on-loading tasks. All code run within the Quantum Inspire platform is run unsupervised. On the other hand, the benefit of this setting is the inherent speed-up for high frequency coupling compared to running user code on an off-site system and off-loading the quantum circuit via the API.

The setting described above, also assumes a specific description of the hybrid tasks. This requires the users to adhere to a certain structure and programming language for the task manager (dispatcher) to be able to orchestrate the classical and quantum runtimes. Quantum Inspire's first approach was a lifecycle scheme. The user would specify an initialize, execution and finalize handle which would be invoked by the task manager. Support structures like advanced error recovery could be more efficiently implemented. However, since most algorithms were based on optimization algorithms, for example from scipy [34], this method proved impractical in the end.

Another interesting point is that with increasing coupling frequency and quantum resources needs, the scheduling of jobs in the quantum backend (job manager) might have to include prioritization and reservation possibilities.

Prioritization will reduce transparency about one's place in the queue but for complex workflows with external resources this will limit the total idle and queueing time on both systems. Potentially, if HPC users would dominate the queue and prevent other users from using the system, fair use policies will have to be implemented. At the current state of the system, this is not yet needed. Reservations could be very useful when the HPC is used as the classical runtime, either due to the size of classical resources needed or, for example, because more control of the classical environment is required. Blocking the quantum resources within a reservation for the entire runtime of the hybrid application is easy, but it can also result in a lot of idling.

## VI. FUTURE PLANS

To better support the continued integration of quantum applications and complex workflow, Quantum Inspire plans to invest in both the end user side and the efficient handling of applications. HPC systems use public information now to determine whether a quantum computer is available. By creating a tighter loopback between the job manager and SLURM/PBS, these systems might better predict availability of the computer, which minimizes the queueing time.

Furthermore, different queueing strategies will be implemented. On the one hand, these ensure that high-frequency tasks are given priority, minimizing the wall clock time for the overall application. On the other hand, these strategies should prevent any HPC from claiming all resources and ensure a fair use. To enable transparency, someone's place in the queue and recent prioritized activity might be conveyed to the end users.

To achieve a speedup within the Quantum Inspire system, the various runtimes might be spun up before a job is created. This would mean idle resources on the Quantum Inspire infrastructure. However, it would almost negate the initialization step for the classical part of a quantum task. This task should in principle employ a hybrid flow, regularly switching between classical and quantum resources. However, in a scenario with low iterations per task, but many tasks as part of a high-level workflow, the initialization penalty might become substantial.

Finally, within Europe there are multiple HPC centers and various upcoming quantum compute platforms [35]. Both, SURF and Quantum Inspire aspire to connect to more systems and become part of the European Infrastructure. We hope that these first integration efforts, experiences and exercises can help guide and develop the European infrastructure.

## VII. CONCLUSIONS

SURF and Quantum Inspire have partnered up together to explore and develop a hybrid classical-quantum infrastructure. Some initial steps include the co-location of a SURF HPC node as a classical runtime and the development of a software stack that allows for the execution of hybrid tasks.

Together with the possibility to execute hybrid tasks, the integration of Quantum Inspire and SURF in the long term aims to enable the execution of hybrid applications. For a hybrid workflow to be executed, the job manager at SURF (SLURM) needs to coordinate with Quantum Inspire. An initial setup uses 2-SLURM clusters to optimize the communication and allow for independent control of the tasks. Although the setup is successful, we are actively testing new potential settings. This setup can easily be extended to allow other HPC centers to connect and use the Quantum Inspire backends.

With these combined efforts, SURF and Quantum Inspire will be able to provide a platform for distributed HPC-quantum computing for the Dutch researchers.

ACKNOWLEDGMENT

This project has received funding from the National Growthfund Quantum Technology KAT-1 program and thanks SURF for funding contribution.

QuTech is a mission driven cooperation between the Technical University Delft and the Netherlands Organization for Applied Scientific Research (TNO).

REFERENCES

[1] A. Montanaro, Quantum Information volume 2, Article number:15023 (2016)
[2] H. P Cheng and E. Deumens and J.K. Freeicks and C.Li and B.A. Sanders, Application of Quantum Computing to Biochemical Systems: A Look to the Future, Phys Chem Chem, 2020, 8, https://doi.org/10.3389/fchem.2020.587143
[3] M. Ruefenacht, *et al.*. Bringing quantum acceleration to supercomputers, IQM, 2022, https://meetiqm.com/uploads/documents/IQM_HPC-QC-Integration-Whitepaper.pdf
[4] P. W. Shor, Scheme for reducing decoherence in quantum computer memory, Phys. Rev. A, 52, R2493—R249
[5] C. Ryan-Anderson, *et al*. Realization of Real-Time Fault-Tolerant Quantum Error Correction, Phys.Rev.X, 11, 041058.
[6] F. Leymann and J. Barzen, 2021, Hybrid Quantum Applications Need Two Orchestrations in Superposition: A Software Architecture Perspective, https://arxiv.org/pdf/2103.04320.pdf
[7] Weder B. and Barzen J. and Leynmann F. and Zimmerman, M., Hybrid Quantum Software Application Need Two orchestrations in Superposition: A Software Architecture Perspective, IEEE, 2021, 1-13
[8] M. P. Johansson and E. Krishnasamy, E. and N. Meyer and C. Piechurski,, Quantum Computing – A European Perspective,2021, https://prace-ri.eu/wp-content/uploads/TR-Quantum-Computing-A-European-Perspective.pdf
[9] T. Peng, A. Harrow, M. Ozols, and X. Wu (2019) "Simulating Large Quantum Circuits on a Small Quantum Computer". (arXiv)
[10] Y. Du and T. Huang and S. You. *et al.* Quantum circuit architecture search for variational quantum algorithms. *npj Quantum Inf* **8**, 62 (2022). https://doi.org/10.1038/s41534-022-00570-y
[11] M. Cerezo and A. Arrasmith, A. and R. Babbush, *et al.* Variational quantum algorithms. *Nat Rev Phys* **3**, 625–644 (2021). https://doi.org/10.1038/s42254-021-00348-9
[12] K. J. Mesman, and F. Battistel and E. Reehuis and D. de Jong, and M.J. Tiggelman amd J. Gloudemans and J. van Oven, and C.C., arXiv, https://arxiv.org/pdf/2303.01450.pdf
[13] T. Lubinski *et al.*, Advancing hybrid quantum-classical computation with real-time execution, Front. Phys, 10, 2022
[14] https://slurm.schedmd.com/documentation.html
[15] https://www.openpbs.org/
[16] https://flux.ly/
[17] https://it4innovations.github.io/hyperqueue/stable/
[18] https://qcg-pilotjob.readthedocs.io/en/develop/
[19] https://research.ibm.com/publications/qiskit-runtime-a-quantum-classical-execution-platform-for-cloud-accessible-quantum-computers
[20] https://github.com/aws/amazon-braket-examples/blob/main/examples/hybrid_jobs/0_Creating_your_first_Hybrid_Job/Creating_your_first_Hybrid_Job.ipynb
[21] Y. Yang, et al, FPGA-based electronic system for the control and readout of superconducting quantum processors, Review of Scientific Instruments, 2022, 93, 074701
[22] https://quantum-computing.ibm.com/lab/docs/iql/manage/systems/reservations
[23] https://www.quantum-inspire.com/
[24] N. Khammassi, I. Ashraf, X. Fu, C. G. Almudever and K. Bertels, ,"QX: A high-performance quantum computer simulation platform," *Design, Automation & Test in Europe Conference & Exhibition (DATE), 2017*, Lausanne, Switzerland, 2017, pp. 464-469, doi: 10.23919/DATE.2017.7927034.
[25] Andre W.Cross and Lev S. Bishop and John A. Smolin and Jay M. Gambetta, OpenQuantum Assembly Language, arXiv, 2017, https://arxiv.org/abs/1707.03429
[26] https://pennylane.ai/
[27] https://qiskit.org/
[28] https://www.surf.nl/en
[29] https://www.quantum-inspire.com/backends/qx-simulator/
[30] https://zeromq.org/
[31] https://zguide.zeromq.org/docs/chapter3/
[32] https://slurm.schedmd.com/rest_api.html
[33] https://www.surf.nl/en/surfdrive-store-and-share-your-files-securely-in-the-cloud/surfdrive-for-users
[34] https://docs.scipy.org/doc/scipy/tutorial/optimize.html
[35] https://eurohpc-ju.europa.eu/selection-six-sites-host-first-european-quantum-computers-2022-10-04_en

APPENDIX A

Simple "ping-pong" implementation between the classical and the quantum runtimes. The example below requires version pyzmq 25.1.0.

*1) Example classical_runtime.py:*

```python
import time
import zmq

context = zmq.Context()
socket = context.socket(zmq.REP)
socket.bind(f"tcp://0.0.0.0:5557")

while True:
    #  Wait for next request from task_manager
    message = socket.recv_string()
    print(f"Generating quantum circuit, based on measurements: {message}")
    time.sleep(1)
    socket.send_string("version 1.0; qubits 2; H q[0]; measure_all")
```

*2) Example quantum_runtime.py:*

```python
import time
import zmq

context = zmq.Context()
socket = context.socket(zmq.REP)
socket.bind(f"tcp://0.0.0.0:5556")

while True:
    #  Wait for next request from task_manager
    message = socket.recv_string()
    print(f"Executing quantum circuit: {message}")
    time.sleep(1)
    socket.send_string('{"001": 512}')
```

*3) Example task_manager.py:*

```python
import zmq

quantum_runtime_port = "5556"
classical_runtime_port = "5557"

context = zmq.Context()
```

```
quantum_socket = context.socket(zmq.REQ)
quantum_socket.connect(f"tcp://localhost:
{quantum_runtime_port}")

classical_socket = context.socket(zmq.REQ)
classical_socket.connect(f"tcp://localhost:
{classical_runtime_port}")

#  Do 10 requests, waiting each time for a response
classical_socket.send_string('{}')
quantum_circuit = classical_socket.recv_string()
print(f"[classical] Received quantum circuit:
{quantum_circuit}")

for request in range(1, 10):
    print(f"Sending request {request}...")

    quantum_socket.send_string(quantum_circuit)
    measurements = quantum_socket.recv_string()
    print(f"[quantum] Received measurements:
{measurements}")

    classical_socket.send_string(measurements)
    quantum_circuit = classical_socket.recv_string()
    print(f"[classical] Received quantum circuit:
{quantum_circuit}")
```

# APPENDIX B

Payload example and code to submit to SLURM's REST API.

*1) Payload example (job.json):*

```
{
  "job": {
    "partition": "p_spin2",
    "tasks": 1,
    "name": "test",
    "nodes": 1,
    "current_working_directory": "/tmp",
    "environment": {
      "PATH": "/bin:/usr/bin/:/usr/local/bin/",
      "LD_LIBRARY_PATH": "/lib/:/lib64/:/usr/local/lib"
    }
  },
  "script": "#!/bin/bash\nsrun hostname\necho 'hello world'\nsleep 300"
}
```

*2) CURL submit code to SLURM REST-API example:*

```
curl -H "Content-Type: application/json" -H X-SLURM-USER-NAME:$QUSER -H X-SLURM-USER-TOKEN:$SLURM_JWT  -X POST http://bquantum.soil.surf.nl:6666/slurm/v0.0.39/job/submit -d@job.json
```